# As-based ternary Janus monolayers for efficient thermoelectric and photocatalytic applications

Poonam Chauhan[1], Jaspreet Singh[1] and Ashok Kumar[1,*]

*[1]Department of Physics, Central University of Punjab, VPO Ghudda, Bathinda, 151401, India*

(April 16, 2023)

*Corresponding Author: ashokphy@cup.edu.in

**Abstract**

Highly efficient and sustainable resources of energy are of great demand today to combat with environmental pollution and the energy crisis. In this work, we have examined the novel 2D Janus AsTeX (X = Cl, Br and I) monolayers using first-principles calculations and explore their potential energy conversion applications. We have demonstrated the thermal, energetic, dynamic and mechanical stability of AsTeX (X = Cl, Br, and I) monolayers. Janus AsTeX (X = Cl, Br and I) monolayers are indirect bandgap semiconductors with high carrier mobilities and excellent visible light optical absorption. Our findings demonstrate that the Janus AsTeCl and AsTeBr monolayers exhibits low lattice thermal conductivity and excellent electronic transport properties obtained using semi-classical Boltzmann's transport theory including various scattering mechanism. Additionally, the redox potential of water is adequately engulfed by the band alignments of the AsTeCl and AsTeBr monolayers. The water splitting process under illumination can proceeds spontaneously on Janus AsTeBr monolayer, while a minimal low external potential (0.26-0.29 eV) is required to trigger water splitting process on Janus AsTeCl monolayer. A more than 10% STH efficiency of these monolayers indicate their potential practical applications in the commercial production of hydrogen. Thus, our study demonstrates that these monolayers can show potential applications in energy conversion fields.

# 1. Introduction

The technologies that produce renewable energies and lessen the energy crisis are widely desired in today's world where environmental pollution and energy crises are two of the biggest problems. Effective solutions to the energy problems include thermoelectric [1,2] and photocatalytic water splitting phenomena.[3,4] The thermoelectric phenomenon directly converts untapped heat into electricity [5,6] while, photocatalytic water splitting uses solar energy to split water into hydrogen and oxygen and transform solar energy into chemical energy.[7] The thermoelectric performance of materials is measured by the dimensionless quantity ZT (figure of merit), whereas, the efficiency for water splitting depends on several factors including proper bands alignments, carrier migration distance, good optical adsorption, and a large number of active sites.[8]

Apparently, the thermoelectric performance of materials can be increased by two ways: improving the power factor (PF = $S^2\sigma$) and decreasing the thermal conductivity $\kappa$ (= $\kappa_e+\kappa_l$). To this end, various ways have been explored to identify novel materials that quantify low lattice thermal conductivity and high electrical conductivity.[9] The thermoelectric performance of two-dimensional (2D) monolayers has been demonstrated to be superior to that of bulk materials due to the distinct scattering mechanism and the energy dependence of the electronic density of states.[10] The thermoelectric performance of graphene, the first ever 2D material, has been reported to be extremely low (ZT < 0.02 at 300 K) due to the high lattice thermal conductivity (>2000 Wm$^{-1}$K$^{-1}$) and poor Seebeck coefficient (< 100 µV/K).[11] The most studied transition metal dichalcogenides (TMDs) exhibits low thermoelectric performance e.g. ZT ~ 0.1 and ZT ~ 0.7 corresponds to $MoSe_2$ and $WSe_2$, respectively at room temperature, which mainly arises due to the high value of lattice thermal conductivity .[12]

Also, due to their exceptional qualities, such as high carrier mobility and good absorption characteristics, 2D materials are incredibly effective photocatalysts for water splitting.[13] Because of high charge transfer efficiency, high optical absorption efficiency, and high stability, TMDs serve as an effective photocatalyst for the process of water splitting.[14, 15] Many TMDs monolayers such as, $MoS_2$, $MoSe_2$[16] and $HfS_2$[17] have been explored theoretically as water splitting photocatalysts. Furthermore, MXenes (transition metal carbides/nitrides) monolayers are also explored as a photocatalyst for water splitting.[18] $Zr_2Co_2$,[19] $Hf_2Co_2$[19] and $Sc_2Co_2$[20] monolayers act as fascinating photocatalysts due to the in-build electric field found in the perpendicular direction which emerges from the non-symmetric structure.[21]

Due to the unique features compared to more traditional 2D materials, a new derivative of 2D materials known as Janus materials recently gained significant scientific attention.[22] Janus materials contain two different types of atoms on different surfaces.[23] Due to out-of-plane asymmetry in Janus monolayers, an intrinsic electric field is induced. After the experimental synthesis of Janus graphene,[24] Zhang and co-workers synthesized the Janus MoSSe[25], while Pan and co-workers synthesized Janus WSSe.[26] The experimental synthesis of these monolayers inspired researchers to further explore the Janus monolayers. Different kind of Janus monolayers such as SbTeI,[27] $In_2SSe$,[28] TiXY (X≠Y; X, Y = Cl, Br and I),[29] $MX_2Y$ (X≠Y=S, Se),[30] MXY (M = Ge, Sn; X≠Y = S, Se),[31] WXO (X = S, Se and Te),[32] $XSn_2Y$ (X≠Y = P, As, Sb and Bi)[33] and MoSSe[34] are also explored theoretically. Various ternary Janus monolayers such as WSTe (ZT~0.74-2.56),[35] BiTeBr (ZT ~ 0.7-1.78),[36] BiTeCl (ZT ~ 0.43-0.75),[37] AsSBr (ZT ~ 0.21-0.91),[38] $AsSbC_3$ (ZT ~ 0.73-0.95),[39] PdSeTe (ZT~0.34-1.86)[40] and PtSeTe (ZT ~ 0.37-1.97)[40] show good thermoelectric performance at the temperature range 300K-800K. The intrinsic electric field induced due to structural asymmetry is favorable to hinder the carrier recombination which enhances the water splitting performance of photocatalysts. Additionally, 2D Janus materials offered different atomic species to execute hydrogen evolution reaction (HER) and oxygen evolution reaction (OER). Various 2D Janus monolayers such as WSSe,[41] MoSSe,[42] WSeTe[43] and PtSSe[44] exhibit high solar-to-hydrogen conversion efficiency.

Recently, the experimental synthesis of Janus BiTeCl, BiTeBr[45] and BiTeI[46] monolayers has inspired researchers to explore more analogous Janus monolayers of this family. Theoretical calculations show that 2D BiTeX (X = Cl, Br and I) Janus monolayers are indirect bandgap semiconductors.[47] BiTeCl and BiTeI monolayers exhibit high value of carrier mobilities of the order of $10^3$ cm$^2$V$^{-1}$s$^{-1}$.[48] BiTeCl and BiTeBr monolayers show appreciable performance in thermoelectric field with figure of merit 0.81[37] and 1.75[36] at 700K, respectively. Note that the ZT value of experimentally synthesized Bulk BiTeX (X=Cl, Br, I)[49][50] at room temperature ranges from ~0.1 to ~0.2. Further BiXY (X = S, Se, Te; Y = F, Cl, Br and I) monolayers shows application in piezoelectric field.[47] Very recently arsenic-based ternary Janus monolayers are also explored in the fields of photocatalytic water splitting using first-principles method.[51] Note that the parent materials of these monolayers are Bi$_2$Te$_3$ and As$_2$Te$_3$.[52] In this work, we have explored the thermoelectric and photocatalytic performance of Janus AsTeX (X = Cl, Br and I) monolayers by employing a first-principles method. Thermoelectric performance of these monolayers is evaluated by including various scattering models' effects (acoustic scattering, optical scattering, polar optical scattering, impurity scattering and piezoelectric scattering) on the electronic transport properties. AsTeCl and AsTeBr monolayers shows indirect band gap with suitable band edge positions and the potential to meet the requirements for redox potentials of water for photocatalysis with more than 10% solar to hydrogen conversion efficiency.

**2. Computational Details**

Quantum ESPRESSO package is used to perform the first-principles density functional theory (DFT)-based calculations.[53] For the treatment of electron-ion interactions and exchange-correlation energies, the norm-conserving pseudopotentials with generalized gradient approximation (GGA) are implemented. The k-point grid of 24x24x1 with plane-wave cut off of 90 Ry is used for the sampling of the Brillouin zone for both structural relaxation and electronic structure calculations. To eliminate the unphysical forces between the adjacent layers, a vacuum of ~ 17 Å is used. The atomic positions and structural parameters are relaxed up to the energy and forces convergence of $10^{-7}$ eV and $10^{-6}$ eV/Å$^2$, respectively. The dipole corrections have been employed to take care of intrinsic electric field to calculate the electrostatic potential difference profiles. To demonstrate the dynamic stability, the phonon dispersion spectra calculations were carried out using density functional perturbation theory (DFPT) with the q-point mesh of 16x16x1 and convergence threshold of $10^{-18}$ Ry. The MD simulations were performed using the Nose-Hoover thermostat algorithm at 300K and 500K for a total of 5ps with a time step of 3fs with relatively large supercell size of 4x4x1 and the k-point mesh of 12x12x1. The electronic properties are calculated using most precise HSE06 (Heyd-Scuseria-Ernzerhof) hybrid functional.[54] To investigate the optical properties involving the excitonic effects, YAMBO code[55] integrated with Quantum ESPRESSO package is used. The plasmon-pole approximation is used to evaluate the quasi-particle energy within $G_0W_0$ method. Bethe-Salpeter equation (BSE) is solved to evaluate the dielectric response of the system involving the electron-hole interactions.

The electronic transport properties usually calculate from constant relaxation time approach do not include any kind of scattering effects and hence, the values come out are not consistent with the experimental results. So, in order to get the consistency with experimental results, we have employed a more accurate approach that includes the effect of various energy and temperature dependent scattering models in the calculations of electronic transport properties as implemented in the PAOFLOW 2.0 package.[56, 57] The method described by Jacoboni et. al[58] and recently incorporated in the BoltzTraP framework by Fiorentini et. al.[59] is used to evaluate the universal relaxation time as:

$$\frac{1}{\tau_{total}(E,T)} = \frac{1}{\tau_A(E,T)} + \frac{1}{\tau_{Op}(E,T)} + \frac{1}{\tau_{POp}(E,T)} + \frac{1}{\tau_I(E,T)} + \frac{1}{\tau_P(E,T)} \quad (1)$$

where, $\tau_A$, $\tau_{Op}$, $\tau_{POp}$, $\tau_I$ and $\tau_P$ terms indicate the acoustic scattering, optical scattering, polar optical scattering, impurity scattering and piezoelectric scattering, respectively. The details of the terms involved in the above equation are given in ESI. The Phono3py code with relaxation time approximation is used to calculate lattice thermal conductivity.[60] A highly optimized mesh cutoff 50x50x1 is used in the Phono3py code. For 2D monolayers, the thermoelectric parameters are normalized by $L_z/d_0$ parameter, where $L_z$ is the length of unit cell in z direction and $d_0$ is the effective thickness which is taken as the sum of van der Waals radii of surface atoms (Te, Cl, Br and I) and thickness of monolayers.

The solvation effect in the aqueous solution is taken under consideration with Poisson-Boltzmann implicit solvation model for photocatalytic calculations. The dielectric constant of liquid water is taken as 78.3.[61] The hydrogen evolution reaction (HER) and oxygen evolution reaction (OER) performances are evaluated by calculating Gibbs free energy (ΔG) given as:

$$\Delta G = \Delta E + \Delta E_{ZPE} - T\Delta S + \Delta G_U + \Delta G_{pH} \quad (2)$$

Here ΔE is the adsorption energy, $\Delta E_{ZPE}$ and TΔS are the difference in zero point energy and entropy, respectively. $\Delta G_U$ (= −eU, U is electrode potential relative to standard hydrogen electrode) stands for extra potential of photogenerated charge carriers (electrons and holes). $\Delta G_{pH}$ (= $K_BT \times \ln10 \times$ pH) is the contribution of Gibbs free energy at different pH concentrations. The more details of the calculations of Gibbs free energy, thermodynamic oxidation and reduction potentials are given in ESI.

## 3. Results and Discussion

### 3.1. Structure, stability and lattice thermal conductivity

The optimized structure of Janus AsTeX (X = Cl, Br and I) monolayers is presented in Fig. 1. These monolayers possess trigonal structure with hexagonal unit cell containing three atoms per primitive cell. The crystal structure of Janus AsTeX (X = Cl, Br and I) monolayers is asymmetric in which As atom is sandwiched in between Te and X atoms, that leads to the electrostatic potential difference of 1.46 eV, 1.30 eV and 0.23 eV for Janus AsTeCl, AsTeBr and AsTeI monolayers (Fig. S1, ESI). Due to difference in the electronegativity of the surface atom an electric field is induced pointed from Te to X atoms. The calculated value of intrinsic electric field is 3.60 eV/Å, 2.32 eV/Å and 0.59 eV/Å for AsTeCl, AsTeBr and AsTeI monolayers, respectively. The relaxed lattice constant and other structural parameters (h, α and β) of AsTeX (X = Cl, Br and I) monolayers are listed in Table 1. The comparison of relaxed parameters of AsTeX (X = Cl, Br and I) with their parent monolayer $As_2Te_3$ is given in Table S1, ESI.

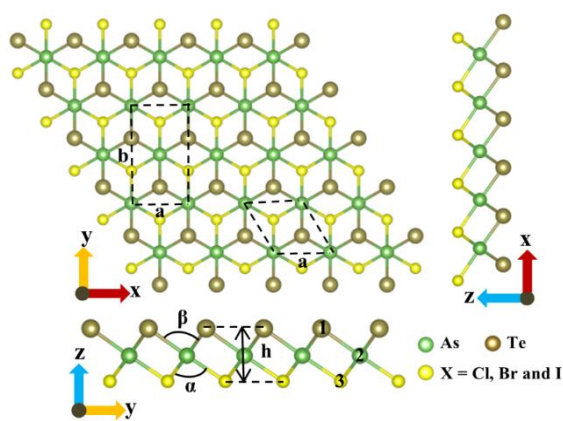

**Fig. 1** Top and side views of the crystal structure of AsTeX (X = Cl, Br and I) monolayers. The hexagonal and rectangular unit cells are shown by dotted line in the top view.

**Table 1** The relaxed parameters: lattice constants (a, b), thickness (h), bond angle (α for Te-As-Te, β for X-As-X) and formation energy ($E_f$).

| 2D Monolayer | (a, b) (Å) | H (Å) | α (°) | β (°) | $E_f$ (eV) |
|---|---|---|---|---|---|
| AsTeCl | 3.93, 6.82 | 3.88 | 91.95 | 89.65 | -5.66 |
| AsTeBr | 3.98, 6.90 | 4.02 | 88.24 | 90.59 | -5.66 |
| AsTeI | 4.09, 7.08 | 4.19 | 84.55 | 92.73 | -5.64 |

**3.1.1 Stability analysis**

To examine the energetic stability of AsTeX (X = Cl, Br and I) monolayers, we calculate their formation energy as:

$$E_f = \frac{[E_{AsTeX} - E_{As} - E_{Te} - E_X]}{3} \quad (3)$$

Here $E_{AsTeX}$ is total energy of AsTeX monolayer unit cell. $E_{As}$, $E_{Te}$ and $E_X$ are the energies of As, Te and X in their bulk stable phases. The calculated value of formation energy is mentioned in Table 1. The negative value of formation energy of AsTeX (X = Cl Br and I) indicates the energetic stability of these monolayers.

Next, the dynamic stability of AsTeX (X = Cl, Br and I) monolayers is assessed by inspecting the phonon spectra of these monolayers corresponding to high symmetry points in the first Brillouin zone as shown in Fig. 2. These monolayers are kinetically stable since there are no discernible imaginary modes in the phonon spectra. The phonon spectrum contains 9 modes of vibrations; 3 acoustic modes (flexural acoustic mode (ZA), transverse acoustic mode (TA) and longitudinal acoustic mode (LA)) and 6 optical modes.

The flexural acoustic mode of AsTeX (X = Cl, Br and I) monolayers exhibits quadratic nature near the Γ point like other 2D materials.[62, 63] The LA and TA mode of these monolayers indicates linear behavior. The highest vibration frequencies of acoustic modes are calculated to be 85.78 cm$^{-1}$, 76.43 cm$^{-1}$ and 73.63 cm$^{-1}$ for AsTeCl, AsTeBr and AsTeI monolayers, respectively.

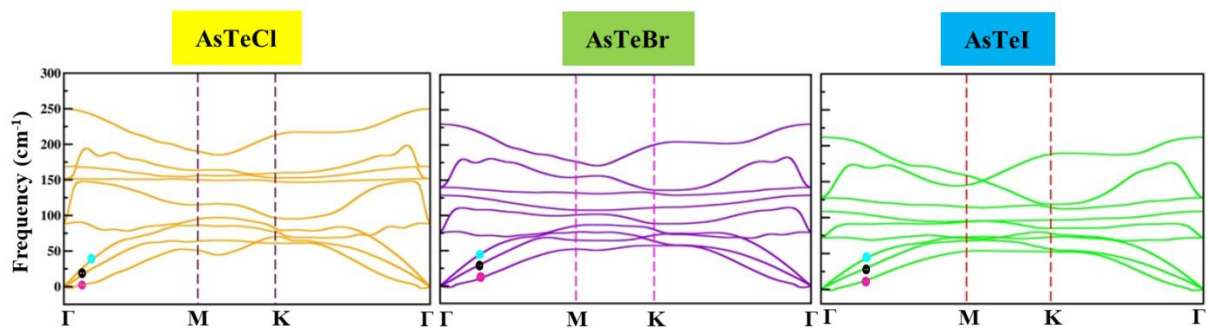

**Fig. 2** Phonon spectra of AsTeX (X = Cl, Br and I) monolayers. Pink, black and cyan balls in low lying phonon branches along Γ-M direction indicates the flexural acoustic (ZA), transverse acoustic (TA) and longitudinal acoustic (LA) modes, respectively.

The frequency of optical modes decreases from AsTeCl to AsTeI, that is associated with the atomic mass of halogen atoms, which increases from Cl to I. According to our computed results, gapless optical and acoustic branches leads to the high scattering process rate and less phonon lifetime, which means that lattice thermal conductivity of these monolayers is relatively low. In all three Janus monolayers, there is a frequency range where both acoustic and optical modes coexist, that may lead to low thermal conductivity due to the possibility of strong optical-acoustic scattering.

Now, the thermal stability of AsTeX (X = Cl, Brand I) monolayers is demonstrated by AIMD simulations with a larger supercell of 4x4x1 size for 5000fs at 300K and 500K (Figure S2, ESI) with 3fs time step. The smaller fluctuations of total energy and temperature around the constant level with time confirm the thermal stability of these monolayers. Our analysis shows that even after 5000 fs, the atomic structures of Janus monolayers are still robust as displayed in Fig. S2, ESI.

Subsequently, stress-strain graphs are employed to evaluate the mechanical stability of these monolayers' (Fig. S3, ESI). The uniaxial strain is applied corresponding to the x- and y-directions in the rectangular unit cell. The axis orthogonal to the stress direction is fully relaxed to confirm that the applied strain is uniaxial. The computed maximum ideal strength of AsTeCl, AsTeBr and AsTeI monolayers is ~10 GPa, ~10 GPa and ~8 GPa at a critical strain limit of 10%, 8% and 10%, respectively along the both x- and y-directions. The same value of critical strain along x- and y- directions is due to the symmetric in-plane unit cell structure of Janus AsTeX (X = Cl, Br and I) monolayers. The ideal strength of AsTeX (X = Cl, Br and I) monolayers is higher than that of other 2D monolayers such as BiTeI (5.97 GPa),[64] β-$Te_2$S (1.16 GPa)[65] and β-$Te_2$Se (7.05 GPa)[65] monolayers. The calculated values of Young's modulus for AsTeCl, AsTeBr and AsTeI monolayers are ~ 87 GPa, ~51 GPa and ~22 GPa, respectively. The lower value of Young's modulus corresponds to the higher flexibility. The calculated value of Poisson ratio for AsTeX (X = Cl, Br and I) monolayers is ~0.2, which is comparable with other Janus monolayers such as BiXY (X = S, Se and Te; Y = F, Cl, Br and I) (~0.2)[47] and SbXY (X = Se, Te; Y = Br, I) (~0.2).[66]

### 3.1.2. Lattice thermal conductivity

Linear Boltzmann transport equation[67] is used to evaluate the lattice thermal conductivity of AsTeX monolayers i.e. given as:

$$\kappa_{l,i} = \sum \sum C_p v_{g,i}^2(\lambda, q) \tau(\lambda, q) \quad (4)$$

where $\kappa_{l,i}$ is the lattice thermal conductivity, $C_p$ is specific heat capacity, $v_{g,i}$ is group velocity, $\lambda$ is the phonon mode and q is the wave vector. The lattice thermal conductivity of these monolayers decreases with temperature which satisfy the temperature dependence relationship as $\kappa_l \propto 1/T$, indicating lattice thermal conductivity is influenced by Umklapp anharmonic phonon-phonon interactions with temperature.[68] The convergence of k-point mesh is done to calculate the lattice thermal conductivity as given in Fig. S4, ESI. The calculated lattice thermal conductivity value at room temperature is 0.92 W/mK, 2.02 W/mK and 3.36 W/mK for AsTeCl, AsTeBr and AsTeI monolayers, respectively (Fig 3). The low value of lattice thermal conductivity of AsTeCl as compared to AsTeBr and AsTeI suggests that AsTeCl shows better thermoelectric performance. The lattice thermal conductivity of AsTeX monolayers is lower than PdSeTe (4.02 W/mK ), PdSTe (5.45 W/mK),[40] and $Al_2SSe$ (3.8 W/mK )[69] and comparable with Janus monolayers such as BiTeCl (1.46 W/mK ),[37] HfSSe (2.2 W/mK),[70] and BiOCl (3 W/mK),[71] respectively. Note that the lattice thermal conductivity of the parent compound of AsTeX monolayers i.e. $As_2Te_3$, is 2.76 W/mK and 1.68 W/mK along x and y directions, respectively.[72]

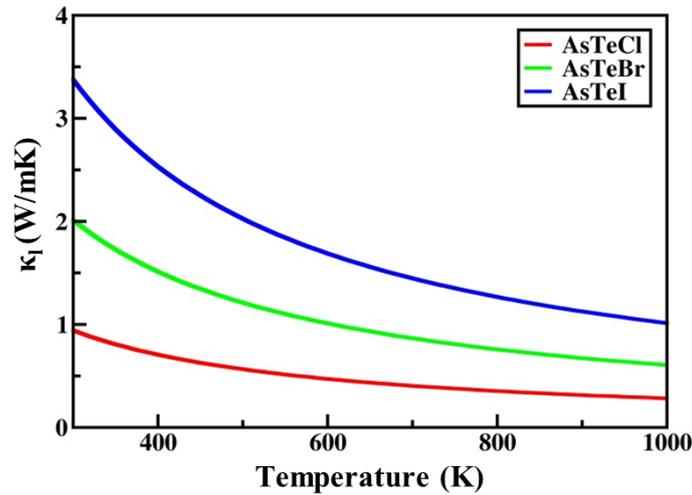

**Fig. 3** Lattice thermal conductivity as a function of temperature for AsTeX (X = Cl, Br, I) monolayers.

We compute phonon group velocity to get a deeper insight into the lattice thermal conductivity[73] as:

$$v_g(\lambda, q) = \frac{\partial \omega(\lambda, q)}{\partial q} \quad (5)$$

where ω(λ, q) is the phonon frequency. The group velocity contribution of the TA and LA mode of AsTeI is higher than that of AsTeCl and AsTeBr, which indicates the higher lattice thermal conductivity of AsTeI monolayer (Fig. S5, ESI). Furthermore, the lattice thermal conductivity is described in terms of a dimensionless quantity, Gruneisen parameter, as:

$$\gamma_{\lambda,q} = \frac{-N}{w_{\lambda,q}} \frac{\partial \omega_{\lambda,q}}{\partial N} \qquad (6)$$

The high value of the Gruneisen parameter is related to the large phonon-phonon scattering, which indicates the lower value of phonon relaxation time. The ZA mode shows the large value of Gruneisen parameter (Fig. S5, ESI) due to the quadratic nature of the band line near the Γ point (Fig. 2). In the case of AsTeI monolayers the value of the Gruneisen parameter is very small which consequences the very high lattice thermal conductivity of this monolayer. The high value of the Gruneisen parameter and the low value of group velocity is favourable factor for the lower value of lattice thermal conductivity of AsTeCl monolayer. Additionally, the phonon lifetime w.r.t. frequency for AsTeX (X = Cl, Br and I) is given in Fig. S6, ESI. The maximum phonon life time's corresponding to AsTeCl, AsTeBr and AsTeI monolayers are ~ 6ps, ~ 8ps and ~16ps, respectively. The phonon life time indicates that AsTeI monolayer possess highest value of lattice thermal conductivity.

**3.2 Electronic Structure and optical absorbance**

Next, the electronic band structures of Janus AsTeX (X = Cl, Br and I) monolayers are evaluated using different levels of theories. Janus AsTeX monolayers are indirect bandgap semiconductors (Fig. 4 and Fig. S7, ESI) with valence band maximum (VBM) located in between Γ and M points and the conduction band minimum (CBM) at Γ point of the Brillouin zone. The calculated value of bandgap using HSE06 method is 2.32 eV, 2.02 eV and 1.78 eV corresponding to AsTeCl, AsTeBr and AsTeI monolayers, respectively. The bandgap values using different levels of theory are given in Table S2, ESI. Due to the heavy atomic mass of the Te atom, we demonstrated the spin-orbit coupling (SOC) effect by calculating band structure using GGA+SOC level of theory. Due to splitting in the valence band, there is small reduction (~0.2 eV) in bandgap value for Janus AsTeX monolayers. Such a small reduction in bandgap values is unlikely to affect the results regarding the thermoelectric and photocatalytic applications, therefore, the computationally expensive calculations including SOC effects considered for further study.

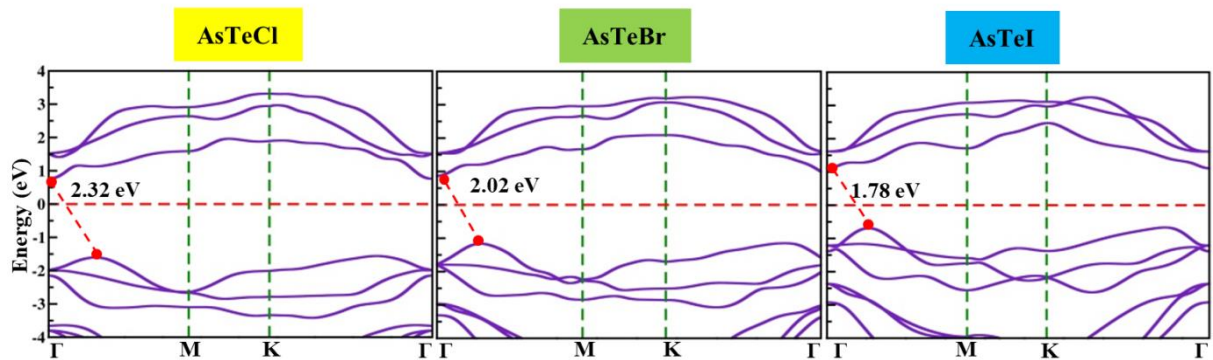

**Fig. 4** The electronic band structure of AsTeX (X=Cl, Br and I) monolayers using the HSE06 hybrid functional. The red balls correspond to the VBM and CBM. The transition between VBM and CBM along with band gap value is also shown. The Fermi level is set at 0 eV.

### 3.2.1. Carrier mobilities

After evaluating the electronic structure of AsTeX (X = Cl, Br and I) monolayers, we now calculated the charge carrier mobility to explore the migration capability of charge carriers (electrons and holes) using deformation potential theory[74, 75]:

$$\mu_c = \frac{e\hbar^3 C^{2D}}{K_B T m m^* E_d^2} \qquad (7)$$

where $\mu_c$ represents the carrier mobility, m is the effective mass, $C^{2D}$ is the elastic modulus and $E_d$ is the deformation potential and these have been evaluated by fitting the curves between energy and strain (Fig. S8, ESI). Our computed results show that the carrier mobility of electrons is one order higher than that of holes. The values of carrier mobility for these monolayers are isotropic in nature as listed in Table S3, ESI. The calculated value of electrons (holes) mobility for AsTeCl, AsTeBr and AsTeI monolayers are ~700 $Cm^2V^{-1}s^{-1}$ (~40 $Cm^2V^{-1}s^{-1}$), ~450 $Cm^2V^{-1}s^{-1}$ (~50 $Cm^2V^{-1}s^{-1}$) and ~300 $Cm^2V^{-1}s^{-1}$ (~39 $Cm^2V^{-1}s^{-1}$), respectively. The higher carrier mobility of electrons arises due to the smaller effective mass and illustrates their high migration potentiality. The carrier mobility of these monolayers are higher than other Janus monolayers such as AsSBr (11.77 $cm^2V^{-1}s^{-1}$),[38] WSSe (229.31 $cm^2V^{-1}s^{-1}$)[35] and WSTe (222.21 $cm^2V^{-1}s^{-1}$)[35]. Note that the carrier mobility of $As_2Te_3$ is calculated to be 183.62 $cm^2V^{-1}s^{-1}$.[76] To further check the mobility anisotropy we use modified longitudinal acoustic phonon limited carrier scattering model. The detailed discussion of carrier mobility calculation using this method is given in ESI. The calculations indicate that carrier mobility using both formulations show no significant difference in the results.

### 3.2.2. Optical Absorbance Spectra

Next, to scrutinize the light-harvesting abilities of AsTeX (X = Cl, Br and I) monolayers, we compute the optical absorbance spectra (A(ω)) using the $G_0W_0$+BSE scheme.[77] The optical absorbance spectra of these monolayers are calculated from the imaginary part of the dielectric function[78, 79] as:

$$A(\omega) = \frac{\omega}{c} L\varepsilon_1(\omega) \quad (8)$$

where L is the measuring parameter for the length of the supercell in the z-direction and $\varepsilon_1$ is the imaginary part of the dielectric function. The first prominent optical absorption peaks for these monolayers lie in the visible region (Fig. 5 (a)). The This indicates the AsTeX (X = Cl, Br and I) monolayers have a good ability for light harvesting.

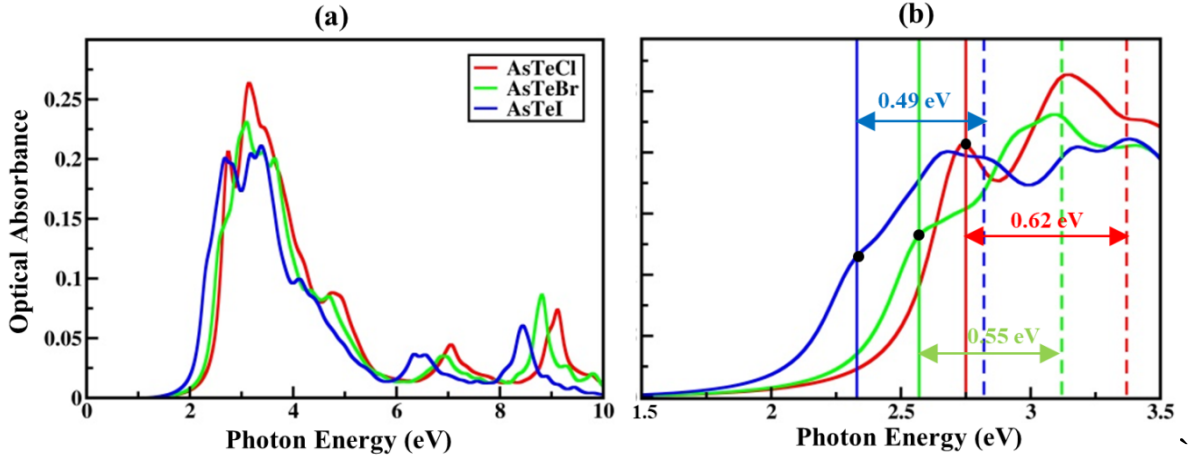

**Fig 5.** (a) Optical absorbance spectra of AsTeX (X= Cl, Br and I) monolayers (b) the same spectra in narrow energy range. The vertical solid lines and dotted lines represent the first excitonic peak and quasiparticle gap, respectively. The exciton binding energy is the difference between first exciton peak and quasiparticle gap. Color code: red-AsTeCl, green-AsTeBr and blue-AsTeI.

We have also calculated the excitonic binding energy as:

$$E_b = E_{QP} - E_{OPT} \qquad (9)$$

where $E_{QP}$ is the quasi-particle direct bandgap and $E_{OPT}$ is the optical bandgap corresponding to the first peak. The GW band structures of all the monolayers are given in Fig. S9, ESI. The calculated excitonic binding energies for AsTeCl, AsTeBr and AsTeI monolayers are 0.62 eV, 0.55 eV and 0.49 eV, respectively as mentioned in Fig. 5(b). The excitonic binding energy of these monolayers is smaller than that of other monolayers such as MgO (2.49 eV), CaO (2.37 eV) [80], BeN$_2$ (1.07 eV)[81] and MoS$_2$ (1.1eV )[82] monolayers, respectively.

### 3.3 Thermoelectric properties of AsTeX (X = Cl & Br) monolayers

Next, we calculate the electronic transport properties to investigate the thermoelectric performance of Janus AsTeX (X = Cl, Br, and I) monolayers. The Seebeck coefficient (S), electronic thermal conductivity ($\kappa_e$), electrical conductivity ($\sigma$), and power factor (P) play vital role for computing the thermoelectric performance (ZT). These parameters are evaluated including various scattering models described in ESI. To assess the electronic transport properties incorporating scattering mechanisms, the semi-classical Boltzmann's equation is used that describes the variation in distribution function *f* under an external electric field. A generating tensor ($\mathcal{L}_\alpha$), α = 0, 1, 2, to describe the transport properties is expressed as[83]:

$$\mathcal{L}\alpha = \frac{1}{4\pi^3} \int \sum_i \tau_i(k) v_i(k) v_i(k) \left(\frac{-\partial f_0}{\partial E}\right) [E_i(k) - \mu]^\alpha dk \qquad (10)$$

where $\tau_n(k)$ is the relaxation time, $v_i(k)v_i(k)$ represent dyadic product, $\varepsilon_n(k)$ the band structure and μ is the chemical potential.

The electronic coefficients S and σ is given as:

$$S = \frac{1}{Te} [\mathcal{L}_0]^{-1} \cdot \mathcal{L}_1 \qquad (11)$$

$$\sigma = e^2 \mathcal{L}_0 \qquad (12)$$

where T is the temperature.

The variation of Seebeck coefficient and electrical conductivity w.r.t temperature is displayed in Fig. 6. The calculated value of Seebeck coefficient for AsTeCl, AsTeBr and AsTeI monolayer is 389.70 µV/K, 219.40 µV/K and 35.12 µV/K (Table S4, ESI), respectively, at 500K temperature. To the best of our knowledge, there is no study that includes the scattering effects in the calculation of electronic transport properties of related 2D materials. So, we have also used the constant relaxation time approach (CRTA) to make comparison with the literature. The calculated values of Seebeck coefficients are 189.1 µV/K , 149.98 µV/K and 71.6 µV/K, respectively at 500K corresponding to which the maximum values of ZT are obtained (Fig. 6(a)). These values are comparable with existing monolayers such as BiTeCl,[37] BiTeBr,[36] WSSe and WSTe[35], respectively. The Seebeck coefficient of $As_2Te_3$ is 475 µV/K and 436 µV/K corresponding to x and y-directions, respectively.[76] The electrical conductivity of AsTeCl, AsTeBr and AsTeI monolayers at 500 K is ~45 ($10^3$/ohm-m), 4.64 ($10^3$/ohm-m) and 1.29 ($10^3$/ohm-m), respectively (Fig. 6(b)). Using constant relaxation time approach, the electrical conductivity of AsTeX monolayers comes out to be of the order of ~$10^6$ (1/ohm.m) which is comparable with other Janus monolayers such as KAgS,[84] KAgSe[84] and BiTeCl.[37]

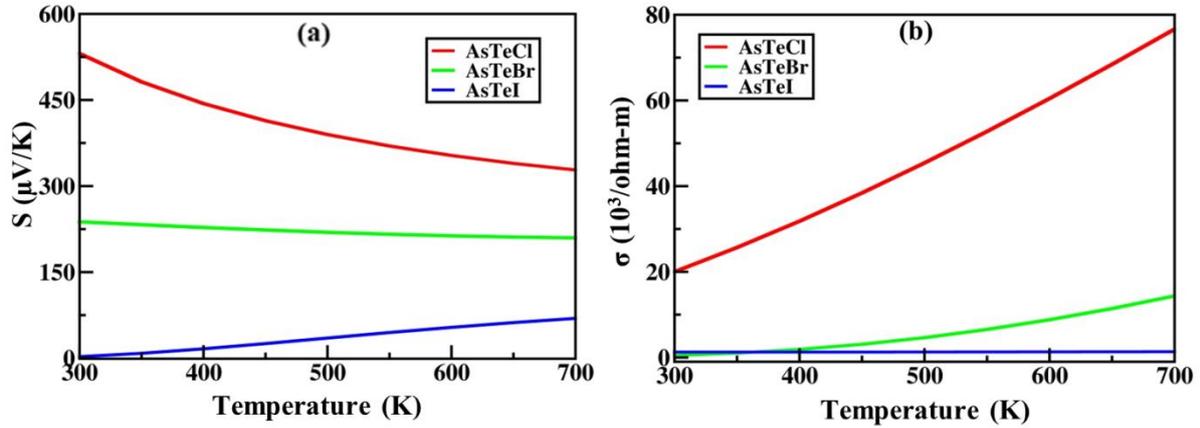

**Fig. 6** The variation of **(a)** Seebeck coefficient (S) and **(b)** electrical conductivity of AsTeX (X = Cl, Br and I) monolayers w.r.t temperature.

The electronic part of the thermal conductivity is obtained by using Wiedemann-Franz law:

$$\kappa_e = L\sigma T \qquad (13)$$

where L is Lorentz number. The electrical thermal conductivity of AsTeX (X = Cl, Br and I) monolayers lies in the range of 5.81-64.75 W/mK while using constant relaxation time approach, these values comes out in the range of 0.18-6.89 W/mK (Table S5, ESI) correspond to the temperature range of 300-500K. Furthermore, the combined effect of electrical conductivity ($\sigma$) and Seebeck coefficient (S) gives the power factor (P) as:

$$P = S^2\sigma \qquad (14)$$

At 500K the power factor of AsTeX monolayers lies in range of 0.0001 - 0.00073 W/mK$^2$. By using constant relaxation time approach, the value of power factor comes out to be 0.003-0.0066 W/mK$^2$ at 500K as given in Table S5, ESI. The detailed discussions of electronic transport properties of these monolayers using constant relaxation time approximation is provided in ESI.

### 3.3.1. Thermoelectric figure of merit (ZT)

By the combined effect of lattice thermal conductivity and power factor, we calculate the ZT of 2D AsTeX (X = Cl, Br and I) monolayers that is given as [85, 86]

$$ZT = \frac{S^2 \sigma T}{\kappa_e + \kappa_l} \tag{15}$$

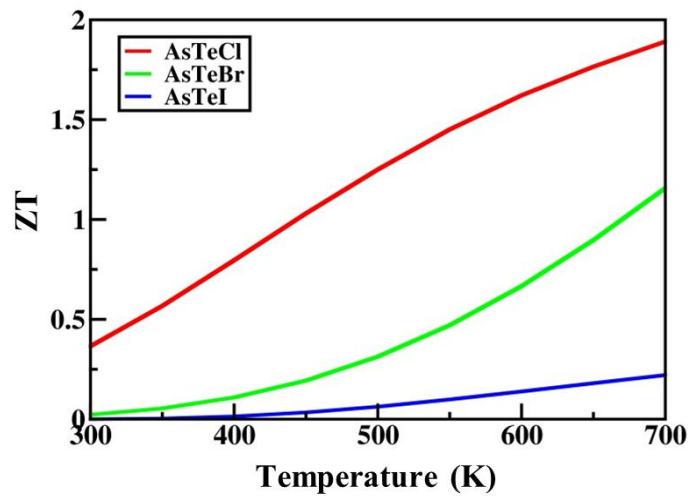

**Fig. 7** Thermoelectric figure of merit (ZT) as a function of temperature for AsTeX (X = Cl, Br and I) monolayers.

ZT as a function of temperature (Fig. 7) indicates the good thermoelectric performance with value of 1.25, 0.31 and 0.06 for AsTeCl, AsTeBr and AsTeI monolayers at 500K (Fig. 7 and Table S6, ESI). By employing constant relaxation time approach, the calculated values of ZT for AsTeCl, AsTeBr and AsTeI monolayers are 3.59, 3.48 and 0.88 (Fig. S11, ESI). These values of thermoelectric figure of merit are higher than MXY (M = Pd, Pt;X,Y = S,Se,Te) (0.33-1.97),[40] SnSSe (3),[87] WSTe (2.56),[35] and comparable with KAgS (4.05)[84] respectively. For the comparison purpose, the thermoelectric figure of merit for other 2D monolayers is listed in Table S7, ESI.

## 3.4. Photocatalytic properties of AsTeX (X = Cl & Br) monolayers

Next, the photocatalytic properties of AsTeX (X = Cl, Br and I) monolayers are examined. Among these, Janus AsTeCl and AsTeBr monolayers are found to possess suitable band alignments for photocatalytic water splitting process. Note that the band alignments of the parent monolayer $As_2Te_3$ do not matches for photocatalytic water splitting.[88] Janus monolayers exhibit different vacuum levels which help to enable hydrogen evolution reaction (HER) and oxygen evaluation reaction (OER), on different surfaces of Janus monolayers (Fig. 8). In the case of AsTeCl and AsTeBr monolayers' conduction band edge with respect to Te side and valence band edge with respect to Cl/Br side cover the redox potential to avail adequate activity for OER and HER, respectively as displayed in Fig. 8(a).

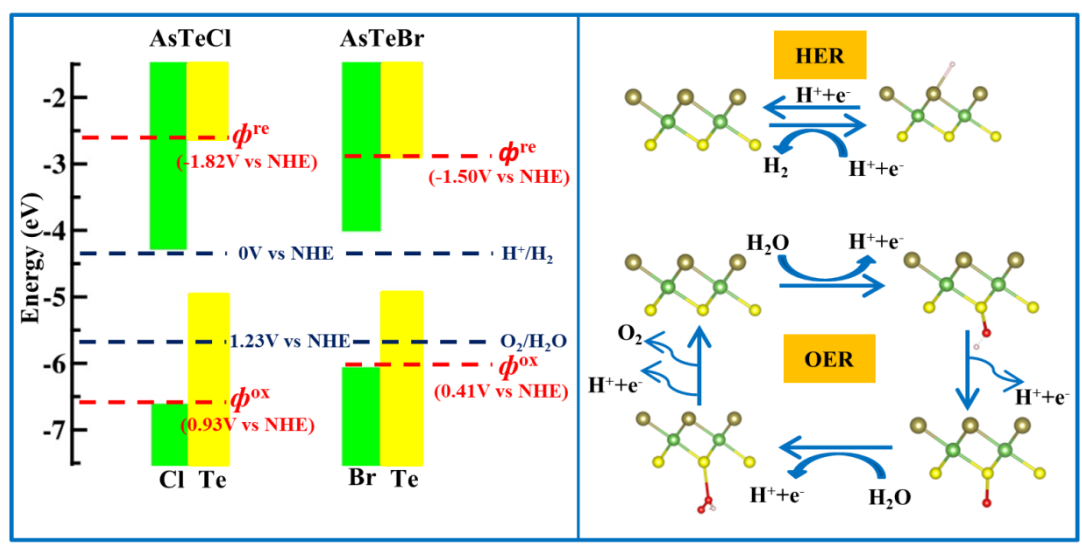

**Fig. 8(a)** Band alignments with respect to vacuum level of AsTeCl and AsTeBr monolayers at HSE06 level of theory. The blue dashed lines specify the water redox potentials at zero pH. The red dashed horizontal lines specify thermodynamic oxidation ($\phi^{ox}$) and reduction ($\phi^{re}$) potentials. **(b)** Proposed photocatalytic pathways and configuration of absorbed species for HER and OER process. Color code: white sphere = H-atom; red sphere= O-atom, Green

sphere = As-atom, Yellow sphere = X (X= Cl, Br)-atoms and Brown sphere = Te atom, respectively).

The stability of the photocatalyst in an aqueous solution is assessed by calculating the thermodynamic oxidation ($\phi^{ox}$) and reduction ($\phi^{re}$) potentials by employing the method of Cheng and Wang.[89] The calculated $\phi^{re}$ and $\phi^{ox}$ are shown by red lines in Fig. 8(a). For both the monolayers, the $\phi^{re}$ ($\phi^{ox}$) is higher (lower) than the reduction potential of $H^+/H_2$ (oxidation potential of $O_2/H_2O$) implies that photogenerated carriers are more likely to reduce and oxidize water molecules than the photocatalyst. Also, the feasibility of adsorption of water molecules on the surface of AsTeCl and AsTeBr monolayers (Fig. S12, ESI) is determined by calculating the adsorption energies of water molecules as:

$$E_{ads} = E_{*H_2O} - E_* - E_{H_2O} \qquad (16)$$

where $E_{*H_2O}$, $E_*$ and $E_{H_2O}$ are the total energies of the adsorbed monolayer, pristine monolayer and water molecule. The calculated value of adsorption energy corresponding to AsTeCl and AsTeBr monolayers are -0.36 eV and -0.32 eV, respectively. The negative adsorption energy of water molecules on these monolayers indicates their feasibility of water molecule's adsorption.

### 3.4.1 Gibbs free energy profiles

We now investigate the Gibbs free energy scheme and mechanisms for half reactions of hydrogen reduction and water oxidation. The calculation details are given in ESI and the various quantities such as zero point energy correction, entropy contribution and total energy is given in Table S8, ESI. For Janus AsTeCl monolayer, the first step involving the combination of proton and electron with monolayer to form the ∗H species in HER mechanism is the rate limiting step (Fig. 9(a, b)). The potential for photogenerated electrons ($U_e$) for HER is evaluated to be 1.5 eV at pH = 0, that lead to the requirement of only 0.26 eV external potential ($\eta_{HER}$) to trigger HER process on AsTeCl monolayer at pH=0. On the other hand, the second step involving the oxidation of $*OH$ to $*O$ species with the release of proton and electron is the rate limiting step in OER process (Fig. 9(c, d)). The potential for photogenerated hole ($U_h$) for OER is 2.05 eV at pH = 7, that lead to the requirement of only 0.29 eV external potential ($\eta_{OER}$) to trigger OER process on AsTeCl monolayer at pH=7. Acidic medium is more favorable for HER while neutral medium is more favourable for OER.

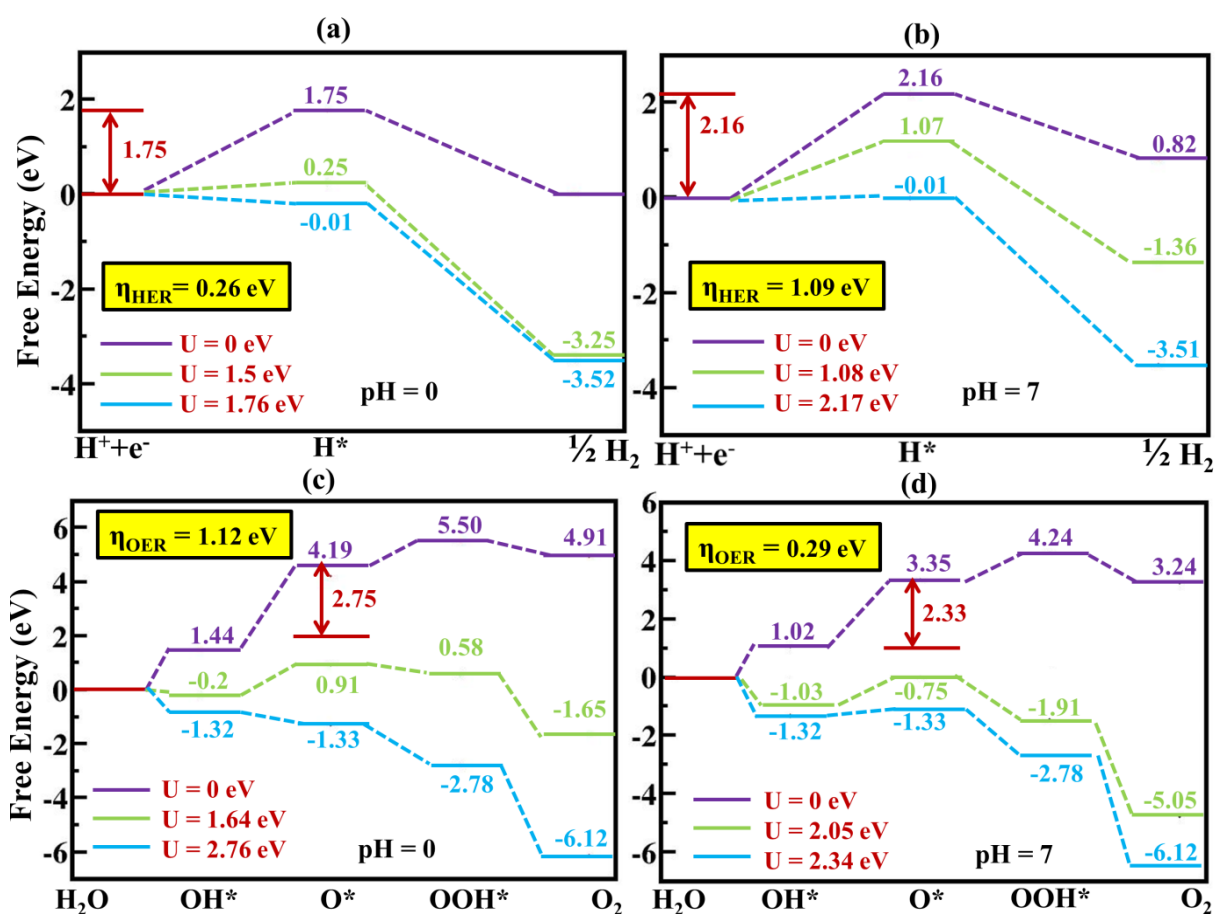

**Fig. 9** Proposed reaction pathways for free energy change corresponding to HER **(a, b)** and OER **(c, d)** for AsTeCl monolayer. The violet and green lines specify the circumstances in the absence and presence of light irradiation (at the potential of photogenerated carriers), respectively, while the sky blue lines represent the situation with required external potentials.

The photogenerated electron (hole) potentials of Janus AsTeBr monolayer is calculated to be 1.5 eV (1.57 eV) and 1.08 eV (1.98 eV) at pH = 0 and pH = 7, respectively that leads to the no requirement of external potential to trigger HER and OER process at pH=0 and pH=7, respectively (Fig. 10). While an external potential of 0.48 eV and 0.76 eV is required for HER and OER process at pH=7 and pH=0, respectively. These values of required external potentials for OER process is less than that of other 2D materials such as GaAs (1.39 eV),[79] β-Te$_2$S (2.01 eV),[65] β-Te$_2$Se (1.30 eV )[65] and CuCl (1.53 eV ).[90] Note that the strain engineering [91] and defect engineering [92] can lower the η$_{OER}$ and η$_{HER}$.

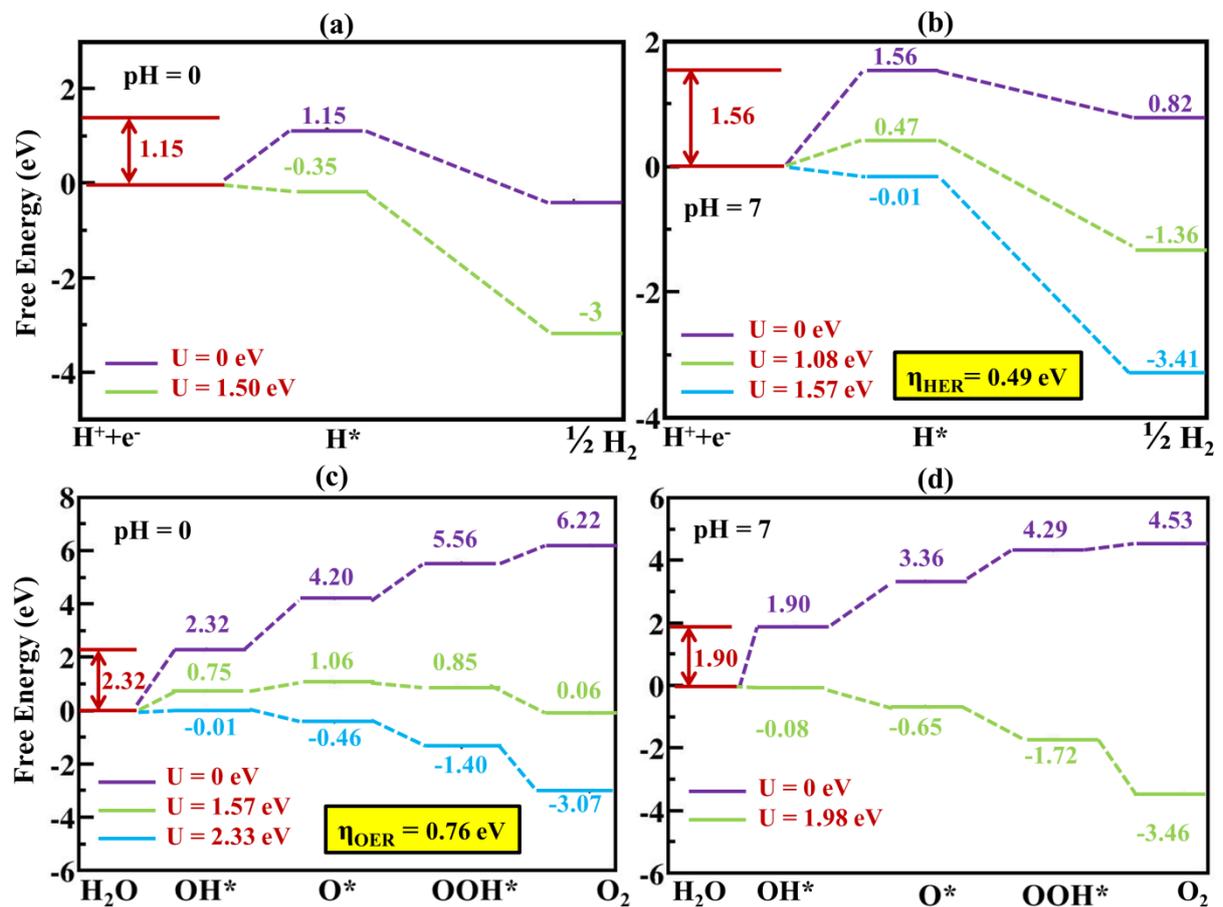

**Fig. 10** Proposed reaction pathways for free energy change corresponding to HER **(a, b)** and OER **(c, d)** for AsTeBr monolayer. The violet and green lines specify the circumstances in the absence and presence of light irradiation (at the potential of photogenerated carriers), respectively, while the sky blue lines represent the situation with required external potentials.

### 3.4.2 Solar-to-hydrogen (STH) efficiency

In AsTeX (X = Cl and Br) monolayers, the proper band alignments and lower external potentials indicating that these monolayers act as potential candidates for desirable photocatalytic characteristics. The ultimate purpose of photocatalytic water splitting using solar energy is to enhance the energy conversion efficiency, which is characterized as solar-to-hydrogen efficiency (STH) that depends on the overpotentials of HER and OER. Based on the proper band alignment, the energy conversion efficiency of AsTeX (X = Cl and Br) monolayers is calculated by using 100% efficiency of catalytic reaction.[93] The computed light absorption ($\eta_{abs}$), carrier utilization ($\eta_{Cu}$) and solar-to-hydrogen efficiency($\eta_{STH}$) are given in Table S10, ESI. The $\eta_{abs}$ depend upon band gap values and carrier utilization efficiencies are connected with χ($O_2$) and χ($H_2$). The STH efficiency of 2D monolayers is given as:

$$\eta_{STH} = \eta_{abs} \times \eta_{Cu} \qquad (17)$$

Corresponding to the maximum value of solar to hydrogen conversion efficiency (Table S10, ESI), the carrier utilization efficiency and light absorption efficiencies for Janus AsTeCl (AsTeBr) monolayers are 53.83% (49.33%) and 35.58% (23.75%), respectively. In the case of Janus monolayer, the presence of an intrinsic electric field that promotes the electrons and holes migration is also considered in terms of potential difference. The corrected efficiency ($\eta'_{STH}$) is illustrated as[65]:

$$\eta'_{STH} = \eta_{STH} \times \frac{\int_0^\infty P(\hbar\omega)d(\hbar\omega)}{\int_0^\infty P(\hbar\omega)d(\hbar\omega) \times \Delta V \int_{E_g}^\infty \frac{P(\hbar\omega)d(\hbar\omega)}{\hbar\omega}} \qquad (18)$$

The calculated corrected solar-to-hydrogen efficiencies values at different pH w.r.t. to χ($H_2$) and χ($O_2$) are given the Table S10, ESI. The maximum corrected solar-to-hydrogen conversion efficiencies for Janus AsTeCl and Janus AsTeBr monolayers are ~ 10% and

~16%, respectively fulfilling the 10% efficiency criteria for cost effective commercial production of hydrogen.[94] The comparison of maximum corrected solar-to-hydrogen efficiencies of AsTeX Janus with other 2D monolayers is listed in Table S11, ESI. Note that the absorbed photon conversion efficiency of WSe$_2$ monolayer is found to be 12% in the experiment of overall water splitting.[95]

## 4. Conclusions

In conclusion, 2D Janus AsTeX (X = Cl, Br and I) monolayers have been explored for their potential applications in the fields of thermoelectric and photocatalytic water splitting. These monolayers exhibit energetic, dynamical, thermal, and mechanical stability. The semiconducting nature of these monolayers is demonstrated by their optoelectronic properties, which include high carrier mobility's and noticeable light absorption abilities. Janus AsTeCl, AsTeBr and AsTeI monolayers exhibit low intrinsic lattice thermal conductivity of 0.92 W/mK, 2.02 W/mK and 3.36 W/mK, respectively at the room temperature. By the combined results of lattice thermal conductivity and electronic transport properties, the calculated value of ZT for AsTeX (X = Cl, Br and I) monolayers lies in between 0.06-1.25 respectively at 500K including different scattering models. At different surface vacuum potentials, the redox potential of water for AsTeCl and AsTeBr monolayers are confined in between the band edge positions. HER and OER can proceeds spontaneously on the surface of Janus AsTeBr monolayer under illumination, while a minimal 0.26 eV and 0.29 eV external potential is required to trigger HER and OER process, respectively on Janus AsTeCl monolayer. The maximum solar-to-hydrogen conversion efficiencies for Janus AsTeCl and AsTeBr monolayers are obtained as ~ 10% and ~16%. Thus, high figure of merit and STH efficiencies of these monolayers indicate their potential applications in energy conversion fields i.e. thermoelectrics and photocatalytic water splitting.

**Acknowledgements**

PC and JS thank the CSIR for providing financial support in the form of a senior research fellowship (SRF). The results presented in this work were obtained using the computational facilities at the department of physics at the Central University of Punjab.